\documentclass[twocolumn,showpacs,preprintnumbers,amsmath,amssymb]{revtex4}
\usepackage{graphicx}
\usepackage{dcolumn}
\usepackage{bm}
\usepackage[scriptsize]{subfigure}

\newcommand{\rmd}{\mathrm d}
\renewcommand{\vec}[1]{\boldsymbol{#1}}

\begin{document}


\title{High-order harmonic generation from field-distorted orbitals}

\author{Maciej Dominik \'Spiewanowski}
\author{Adam Etches}
\author{Lars Bojer Madsen}%
\affiliation{
Department of Physics and Astronomy, Aarhus University, DK-8000 Aarhus C, Denmark
}

\date{\today}

\begin{abstract}
We investigate the effect on high-order harmonic generation 
of the distortion of molecular orbitals by the driving  laser field. Calculations 
for high-order harmonic generation
 including orbital distortion are performed for N$_2$.
Our results allow us to suggest  that field-distortion is the reason why  
the  two-center interference 
minimum has never been observed experimentally in N$_2$. 
We propose experimental parameters which should allow an observation of the two-center 
interference minimum.
\end{abstract}

\maketitle

\section{Introduction}
During the last decades, high-order harmonic generation (HHG) has 
been subject to extensive experimental and theoretical studies. Apart from being
a source of coherent XUV radiation~\cite{Rundquist1998,Bartels2000}, and ultrafast 
pulses~\cite{Drescher2001,Hentschel2001,Sansone2006},
the process of HHG has also been used to extract internuclear 
separations~\cite{Baker2006,Baker2008,li2008,Zhou2008} 
and electronic orbitals~\cite{Itatani2004,Patchkovskii2006,Patchkovskii2007,Haessler2010,Vozzi2011}. 
The intrinsic timing in the 
HHG process associated with the time from ionization to recombination 
can, e.g., be used to resolve nuclear dynamics on an ultrafast timescale~\cite
{Baker2006,Baker2008,li2008}. Turning to orbital tomography, the 
HHG signal has been used to reconstruct molecular orbitals:
the highest occupied molecular orbital (HOMO) in N$_2$~\cite
{Itatani2004,Patchkovskii2006,Patchkovskii2007,Haessler2010}, the HOMO-1
in N$_2$ \cite{Haessler2010}, the HOMO in CO$_2$~\cite{Patchkovskii2007,Vozzi2011} 
and the HOMO in O$_2$~\cite{Patchkovskii2007}. 
Given the experimental laser intensities, it is surprising that, in all of these cases, 
there has been no measurable signature of field-distortion 
in the reconstructed orbitals. The present work identifies for the first time
a clear signature of significant field-induced orbital distortion in the HHG process
by focusing on the effect of the distortion on the two-center interference 
minimum~\cite{LeinPRL2002,LeinPRA2002,Kanai2005,Vozzi2005}.
Indeed, one of the most prominent features of an HHG spectrum is the
presence of a minimum. Such a minimum may have different origin depending on 
the target and laser parameters, but it 
always carries important information about the target.
For example,  a Cooper minimum holds information about the electronic structure of the 
target ~\cite{L'Huillier1993,Farrell2009,McFarland2009,Higuet2011,BertrandPRL2012}. 
Participation of multiple orbitals 
in the HHG process gives rise to a so-called 
dynamic minimum which allows elucidation of the interplay between the  HOMO and 
lower-lying orbitals in the HHG process~\cite{Smirnova2009,Fowe2010}. 
Finally, a two-center interference minimum  gives access to structural information~\cite{LeinPRL2002,LeinPRA2002,Kanai2005,Vozzi2005}. 

Accounting for the effects of orbital distortion allows us to
resolve a long-disputed  puzzle in N$_2$:
Both experimental and theoretical works have proven the existence of Cooper 
minima~\cite{L'Huillier1993,Farrell2009,McFarland2009,WornerPRL2010,Higuet2011,BertrandPRL2012}
and the role of multiple molecular 
orbital contributions in HHG spectra~\cite{Jin2012}. Yet the two-center 
interference minimum, though theoretically predicted~\cite{Zimmermann2005,Odzak2009}, 
has never been observed experimentally~\cite
{Itatani2004,Mairesse2008,BertrandPRL2012}. 
We explain the lack of a two-center interference minimum in terms of laser-induced orbital distortion.
We also show that for specific field parameters 
and orientation of the molecule it should be  possible to observe 
experimentally the two-center minimum in the long trajectories of the HHG spectrum.

\section{Model}
We assume that the 
electronic structure of the molecule adiabatically adjusts to the time-varying external  field, ${\bm F}(t)$. 
With this assumption the Lewenstein model \cite{Lewenstein1994} is extended to include not  
only  the Stark shift~\cite{Etches2010}, but also 
the field-distortion of the initial 
and final states.
For sake of clarity 
we present the approach 
with participation of only the HOMO in the HHG process (the methodology can be 
straightforwardly extended to include more orbitals). The 
HHG spectrum along the polarization
 $\bm{\epsilon}$ can then be calculated as ~\cite{jan}
\begin{equation}\label{spectrum}
S_{\bm{\epsilon}}(\omega)=\left|\int_{-\infty}^\infty 
\bm{\epsilon} \cdot \langle\vec{\hat{v}}(t)\rangle e^{i\omega t}\rmd t\right|^2,
\end{equation}
where the expectation value of the dipole velocity operator reads
\begin{equation}\label{dipole}
\langle\vec{\hat{v}}(t)\rangle=-i\int_{-\infty}^t  \rmd t' \int_{\mathbb{R}^3}\rmd \vec{k} M_\text{rec}(t)e^{-iS(\vec
{k},t,t')}M_\text{ion}(t')+c.c.,
\end{equation}
with
\begin{equation}\label{ion}
M_\text{ion}(t')=\langle\psi_{\vec{k}}^V(t')|V_\text{L}(t')|\psi(\vec{F}(t'))\rangle
\end{equation}
the ionization matrix element,
\begin{equation}\label{phase}
S(\vec{k},t,t')=\int_{t'}^t  \frac{1}{2}\left(\vec{k}+\vec{A}(t'')\right)^2\rmd t''-\int_{t'}^t E(t'')\rmd t''
\end{equation}
the phase accumulated while the electron propagates from time $t'$ to $t$, and 
\begin{equation}\label{rec}
M_\text{rec}(t)=\langle \psi(\vec{F}(t))|{\vec{\hat{v}}}|\psi_{\vec{k}}^\text{V}(t)\rangle,
\end{equation}
the recombination matrix element. The Stark-shifted HOMO energy is denoted by  $ E(t)$ 
in Eq.~\eqref{phase}. Using perturbation theory 
$E(t) = E_0 - {\bm \mu}\cdot {\bm F}(t)  - 1/2 {\bm F}(t)^{\text T}  \underline{\underline{\bm \alpha}}
  {\bm F}(t) $  with $E_0$ the field-free HOMO energy, 
  ${\bm \mu}$ the dipole of the HOMO and $\underline{\underline{{\bm \alpha}}}$ the 
 polarizability tensor. In Eqs.~\eqref{ion} and \eqref{rec} the HOMO
 $|\psi(\vec{F}(t))\rangle$ 
depends explicitly on the instantaneous value of the driving laser field $\vec{F}(t)$. 
Propagation of the continuum electron with momentum $\vec{k}$ is described by the Volkov 
state $|\psi_{\vec{k}}^\text{V}(t)\rangle$, and $V_{\text L}(t)= \bm{F}(t) \cdot {\bm r}$ describes the interaction with the laser field.
We obtain initial and final states by calculating field-dependent orbitals with quantum 
chemistry methods~\cite{gamess} using an aug-cc-pTZV 
basis set~\cite{Dunning1989,Dunning1992}. This approach was successfully used 
to study  deformations in CO$_2$ molecules subject to even stronger fields than used here~\cite
{Kono2001}.
To obtain the spectrum, the field-distorted orbitals and corresponding energies are
calculated at every time step needed for an accurate evaluation of 
the integral in Eq.~(\ref{dipole}).
This approach incorporates  many-electron effects on the bound state
since the calculation of molecular orbitals exposed to the laser field involves 
the multielectron response. 
We note that the present approach requires that the distorted orbitals are calculated at the 
instantaneous value of the external field. Hence, the approach refers to a description using 
the length gauge for the electron-laser interaction, which allows the usage of standard 
quantum chemistry program packages for static-field calculations. The HOMO obtained in 
this way may be transferred to the velocity gauge by a suitable unitary
 transformation~\cite{madsen}.

\section{Results}

We freeze the nuclei in N$_2$  at the internuclear distance of
2.08~a.u. and set the ionization potential  to  15.56~eV. 
The molecule is aligned~\cite{stapelfeldt_review} at $\beta= 75^{\circ}$ with respect to the polarization axis of a linearly polarized driving pulse. The pulse has a five-cycle trapezoidal envelope with one optical cycle for ramp-up and -down. The peak intensity is $9 \times 10^{13}$~W/cm$^2$, and the carrier wave length is 1300~nm. For this orientation and
intensity it can be safely assumed that multiple orbital contributions are negligible 
(see Ref.~\cite{Jin2012}) and therefore we 
restrict ourselves to the signal emitted by electrons detached  from 
the HOMO. Possible depletion effects (not treated here) are not expected to affect the shape 
of the HHG spectra, but only the overall yield, and will, hence, not alter our main 
conclusions.

\begin{figure}[t]
\includegraphics[width=0.495\textwidth]{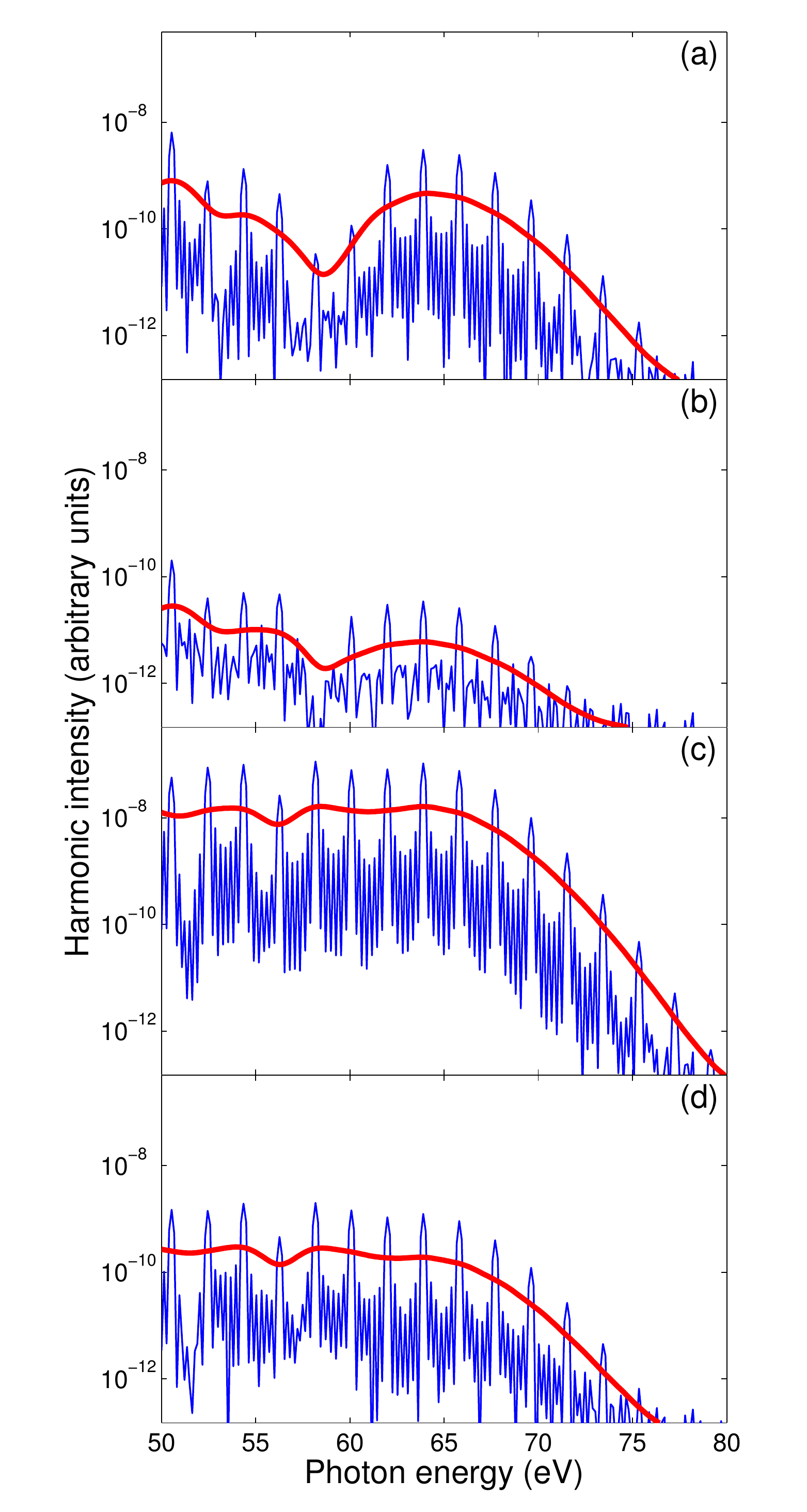}
\caption{(Color online) HHG spectra for N$_2$ (see text for laser parameters and 
orientation). 
The thick (red) curves indicate spectra smoothed with a Gaussian distribution with 
full-width-half-maximum of 4 harmonics. 
(a) Field-free HOMO  in ionization [see Eq.~\eqref{ion}] and recombination [see Eq.~\eqref{rec}] steps. 
(b) Field-distorted HOMO in ionization and field-free HOMO in recombination steps. 
(c) Field-free HOMO in ionization and field-distorted HOMO in recombination steps. 
(d) Field-distorted HOMO in ionization and recombination steps.}
\label{fig:Spectra}
\end{figure}

When field-free HOMOs are used in the ionization [Eq.~\eqref{ion}] and recombination 
[Eq.~\eqref{rec}] steps the model reduces to that of Ref.~\cite{Etches2010}, i.e., the Lewenstein model
~\cite{Lewenstein1994} including Stark shifts. This model predicts a two-center
interference minimum at $\sim58$~eV [Fig.~\ref{fig:Spectra}(a)]. 
The minimum is identified as a two-center interference minimum by its change in 
position when varying the alignment angle: We have checked that the 
minimum moves to higher photon energy when increasing 
the alignment angle~\cite{LeinPRL2002,LeinPRA2002,Odzak2009}.
Such a minimum 
has never been observed experimentally (see, e.g., Refs.~\cite{Itatani2004,Mairesse2008}),
irrespective of orientation and laser intensity. The two-center interference minimum  of concern 
should not be confused with the Cooper minimum, which appears at $\sim38$~eV~\cite
{BertrandPRL2012,WornerPRL2010}. The latter minimum is related to scattering 
properties of the molecular potential~\cite{WornerPRL2009} and is not considered
in the present work, where we focus at larger energies, where the approach treating the 
continuum states as Volkov waves is reasonably accurate.
Next we use a field-distorted HOMO in the ionization matrix element, leaving the state in 
the recombination matrix element field-free. The two-center interference 
minimum remains visible in the spectrum [Fig.~\ref{fig:Spectra}(b)], and hence we reach the 
conclusion that the deformation of the initial state does not significantly obscure the structural 
information associated with the two-center minimum. An important change as compared with 
Fig.~\ref{fig:Spectra}(a) is a much lower yield which is caused by 
destructive interference between amplitudes originating from different ionization times. In the 
next step we examine the role of the recombination matrix element [see Eq.\eqref{rec}] and
use a field-distorted HOMO in the recombination, and a field-free HOMO in the 
ionization step. The result is depicted in Fig.~\ref{fig:Spectra}(c). The disappearance of the 
minimum proves that the recombination matrix element
is sensitive to orbital distortion, and in this sense  holds structural information 
about the target. Similarly, the minimum disappears when taking into account the
field-distortion of the  
HOMO in both the ionization and recombination steps [Fig.~\ref{fig:Spectra}(d)].

\begin{figure}
\includegraphics[width=0.5\textwidth]{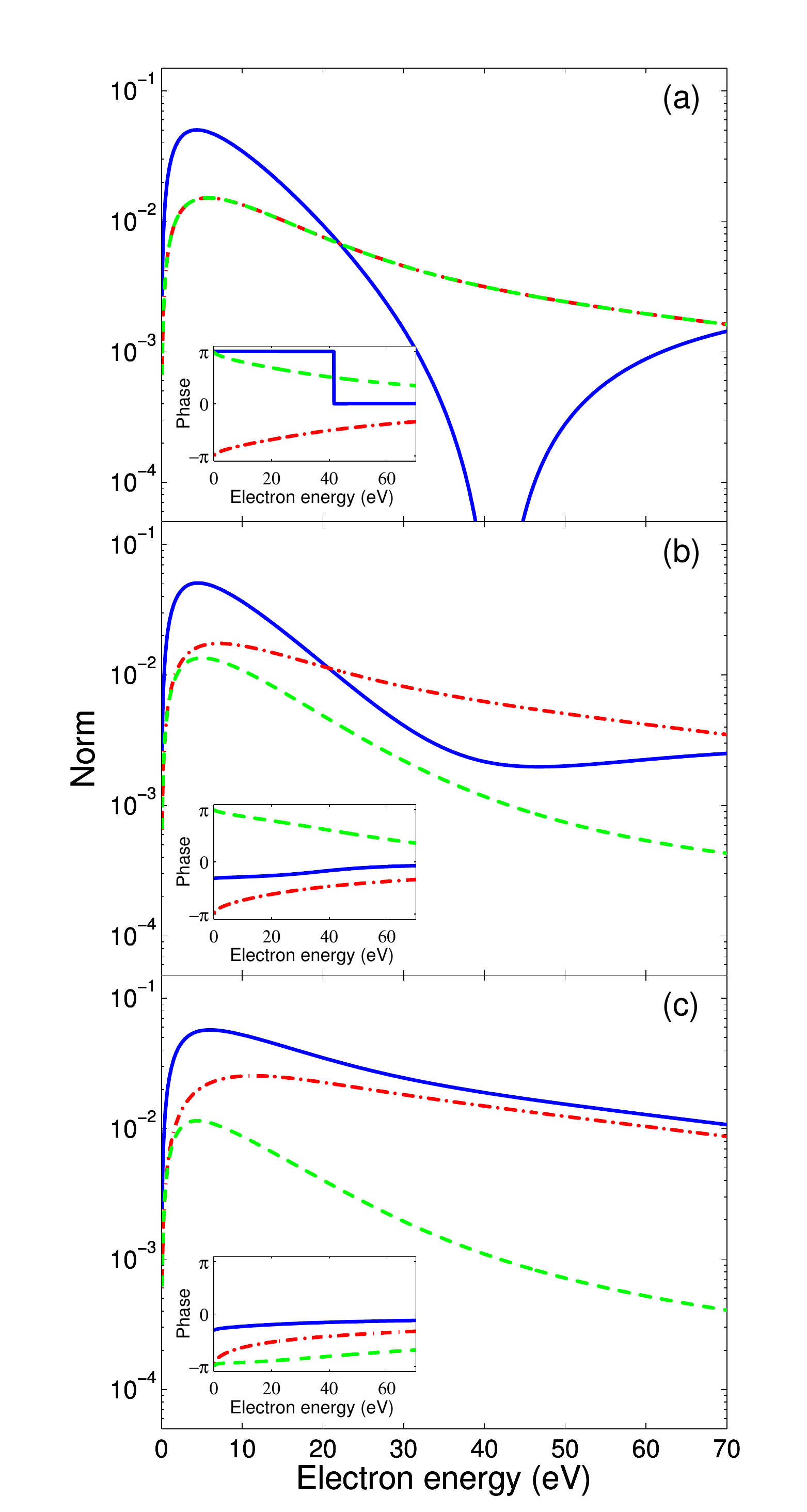}
\caption{(Color online) Norm and phase (inserts) of recombination matrix 
elements [Eq.~\eqref{rec}]
for (a) $F=0$~a.u., (b) $F=0.01$~a.u., (c) $F=0.04$~a.u.
The full (blue) curves show results using the field-distorted HOMO, dot-dashed (red) curves 
denote field-distorted recombination on the N atom at the negative $z$ axis in the molecular 
frame, and  dashed (green) 
denote field-distorted recombination on the N atom at the positive $z$ axis in the molecular 
frame.}
\label{fig:RecombinationMatrixElements}
\end{figure}

The two-center interference minimum appears when the recombination matrix elements 
on both atomic centers are of similar magnitude and  their phase difference is $\pi$. 
To calculate the recombination matrix element on a given center 
we use that the orbitals are expanded in atomic centered Gaussian 
basis functions and take the part of the molecular orbital centered 
on the atomic center of interest \cite{Etches2011}.
In Fig.~\ref{fig:RecombinationMatrixElements} we show how orbital distortion significantly changes the 
relative strengths and phases.
If the HOMO is field-free in the recombination step [Figs.~\ref{fig:Spectra}(a) and (b)], 
then the norms of recombination matrix elements on both centers are equal as shown in
Fig.~\ref{fig:RecombinationMatrixElements}(a). The phase difference reaches $\pi$ for a 
continuum electron energy of  $\sim42$~eV
corresponding, by adding the ionization potential,   to a harmonic energy of $\sim58$~eV. 
This explains the position of the  minima in 
Figs.~\ref{fig:Spectra}(a) and (b). 
When taking into account field-distortion in the recombination step the value of the field 
changes with recombination time. 
In Figs.~\ref
{fig:RecombinationMatrixElements}(b) and (c) we depict the norm and phase of  the
recombination matrix elements when the HOMO is distorted by  $F= 0.01$ ~a.u. and 
$F=0.04$ a.u., respectively. 
These figures show that the norms on different centers become increasingly different for 
increasing field. 
As a consequence the two-center interference minimum weakens and subsequently totally 
vanishes. 
Monitoring the variation of the minimum with the field strength
allows us to specify a range of field values for which observation of the minimum is 
possible. This range 
corresponds to recombination at fields $|F| \lesssim 0.012$~a.u., and we will refer to it as the weak-field region.

\begin{figure*}[t]
\includegraphics[width=\textwidth]{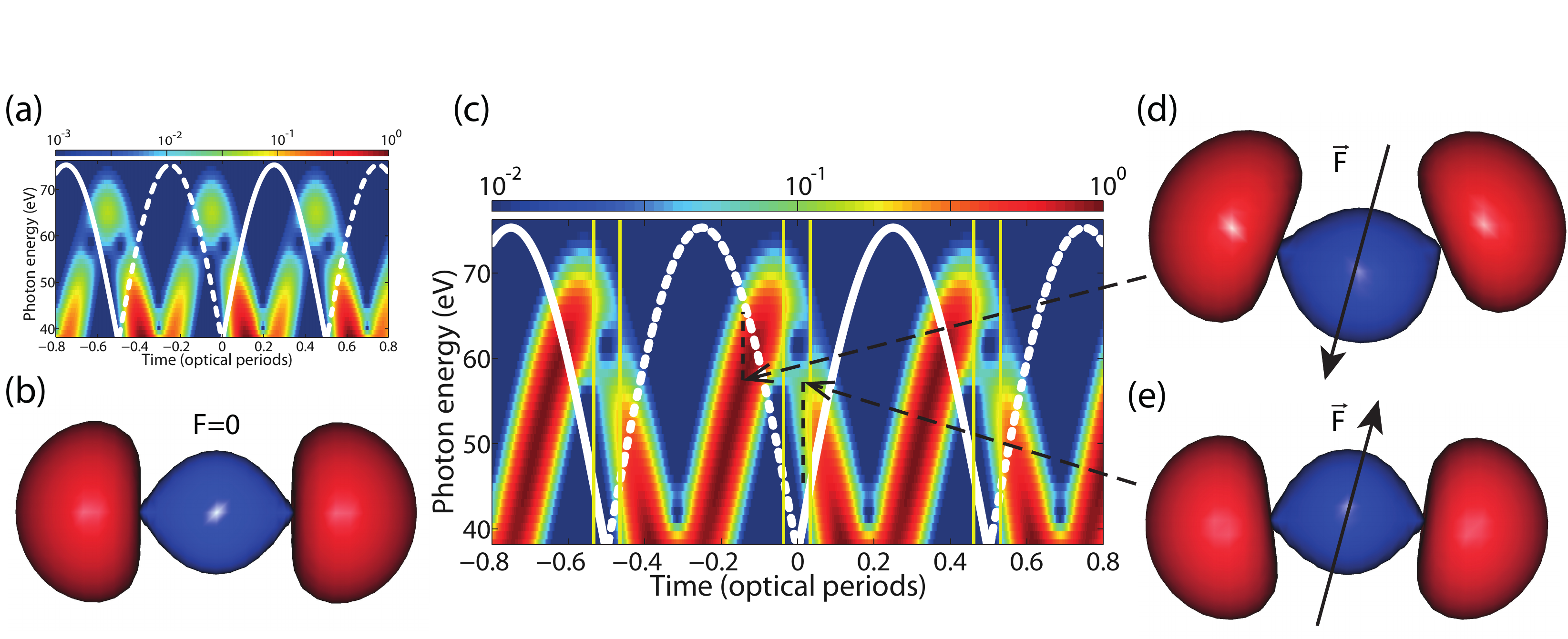}
\caption{(Color online) Emission times of the spectra presented in Figs.~1(a) and 1(d). 
(a) Emission times when performing the calculation with undistorted HOMOs in the ionization and recombination steps
[Fig.~1(a)]. Curves indicate positive (full, white) and 
negative (dashed, white) values of the laser field at the recombination time. 
(b) Field-free, undistorted HOMO.
(c) Emission times when performing the calculation with field-distorted HOMOs in the ionization and recombination steps 
[Fig.~1(d)]. 
Curves indicate positive (full, white) and 
negative (dashed, white) values of the laser field at the recombination time. 
The vertical (yellow) lines identify the weak-field regions (see text).
(d) Field-distorted HOMO for $F=0.04$~a.u.. (e) Field-distorted HOMO for $F=0.01$~a.u.
In (d) and (e) 
full (black) arrows  indicate the field direction. 
The arrows from (d), and (e) to (c) indicate 
the recombination times for which the HOMOs are depicted, 
and dashed (black) lines in (c) guide the eye to the corresponding value of the field.
}
\label{fig:Gabor5cycle}
\end{figure*}

While Fig.~\ref{fig:RecombinationMatrixElements} defines the weak-field region, it does not contain 
information about the value of the field at the time of recombination. 
To this end we consider a time-profile analysis~\cite{tp1,tp2,tp3}.
Figure \ref{fig:Gabor5cycle} presents such an analysis corresponding to the spectra in Figs.~1(a) and 1(d).
The figure shows which recombination, i.e., emission times contribute to the signal at a given photon energy. 
Figures 3(a) and (c) show that the harmonics  come in bursts each half cycle. 
Each harmonic burst consists of two branches, known as 
the short and long trajectories~\cite{gaarde}. 
The short trajectories start less than 2/3 of  an optical cycle before recombination and
give rise to  branches with positive slope. The long trajectories start more than 2/3 of an
optical cycle before recombination, and give rise to branches with negative 
slope.  Figure 3(a) shows that when the calculation is performed using the undistorted HOMO shown in Fig.~3(b), 
there is a minimum at around 58 eV in
both branches, consistent with the minimum in Fig.~1(a). Figure 3(c) shows that 
when field-distortion is included (see Figs.~3(d) and 3(e) for snapshots of the distorted orbitals), 
the short trajectory branches are more intense than the long trajectory branches, and no minima occur in the former.
For example, the short trajectory branches give rise to photons at 61 eV while the 
corresponding long trajectory branches do not contribute significantly to this photon energy. In Fig.~3(c) the vertical (yellow) 
lines indicate the time windows where the field at recombination is within the weak-field region.
We see that the trajectories that recombine in these windows belong to the long trajectory branches. 
The minimum at 61 eV for the long trajectory branch is the two-center interference minimum, 
that is shifted in position by orbital-distortion  from 58 eV in the 
undistorted case [Fig.~3(a)]. Following the discussion in connection with Fig.~2, this implies that the two-center interference 
minimum can only be observed in 
the long trajectory branch, and therefore
a measurement isolating the contribution to the spectrum from  the long trajectories  may lead to an observation of a two-center 
interference minimum in N$_2$. 

The reason why the minimum is not seen in Fig.~\ref{fig:Spectra}(d) is that the short trajectory 
harmonics dominate the spectrum in our calculation (see Fig.~\ref{fig:Gabor5cycle}). To isolate the spectrum originating from the 
long trajectories,
macroscopic phase matching should be used to select
the long trajectories by either changing the focusing conditions of the driving field~\cite{antoine} or by 
far-field separation~\cite{bellini} (see also the review \cite{gaarde} and references therein).

\section{Conclusion and outlook}

Our results indicate that field-distortion of molecular orbitals can be important  for 
understanding spectral features of molecular 
targets. The effect of orbital distortion is manifested in both ionization and 
recombination events.  The orbital distortion in the ionization event  results in a lower 
HHG yield in N$_2$ compared to the undistorted case. 
Orbital distortion in the recombination step influences the structural information
encoded in the spectrum. For example, 
the absence of a two-center interference minimum in the HHG spectrum of
N$_2$ is a clear signature of orbital 
distortion taking place when the short trajectory recombines outside the weak-field limit. 
By carefully adjusting the molecular orientation and the laser parameters, conditions can be obtained 
such that the long trajectory recombines within the weak-field limit, allowing the 
observation of the N$_2$ two-center interference minimum.

The field-distorted orbitals can be obtained in standard quantum
chemistry software packages~\cite{gamess}, and 
therefore the methodology  can be applied to many molecular systems.
The approach opens up the possibility to study a range  of questions
that can shed more light on the process of HHG. For example, the effect of the 
orbital distortion will be very sensitive to the value of the polarizability.
N$_2$ has a relatively high polarizability, and therefore the 
effect on the spectrum is large. In other systems, for example CO$_2$, the polarizability 
is much smaller, and orbital distortion should be less important.
In addition, an approach including orbital distortion 
is highly relevant for orbital tomography. Our work indicates that 
HHG-based orbital tomography will never allow the reconstruction of a field-free orbital, but 
rather gives a coherent average over field-distorted orbitals. 
The degree of field-distortion may be reduced by using short driving pulses~\cite{Zwan2008}, 
and isolating the contribution from the long trajectories.
Finally, we note that the present approach  can also 
be used to address the effect of  orbital-distortion 
in the strong-field approximation 
for ionization~\cite{Keldysh,faisal,reiss}.

\begin{acknowledgments}
This work was supported by 
the Danish Center for Scientific Computation,
the Danish Natural Science Research Council,
and an ERC-StG (Project No.~277767 - TDMET).
\end{acknowledgments}


\begin{thebibliography}{49}
\expandafter\ifx\csname natexlab\endcsname\relax\def\natexlab#1{#1}\fi
\expandafter\ifx\csname bibnamefont\endcsname\relax
  \def\bibnamefont#1{#1}\fi
\expandafter\ifx\csname bibfnamefont\endcsname\relax
  \def\bibfnamefont#1{#1}\fi
\expandafter\ifx\csname citenamefont\endcsname\relax
  \def\citenamefont#1{#1}\fi
\expandafter\ifx\csname url\endcsname\relax
  \def\url#1{\texttt{#1}}\fi
\expandafter\ifx\csname urlprefix\endcsname\relax\def\urlprefix{URL }\fi
\providecommand{\bibinfo}[2]{#2}
\providecommand{\eprint}[2][]{\url{#2}}

\bibitem[{\citenamefont{Rundquist et~al.}(1998)\citenamefont{Rundquist, Durfee,
  Chang, Herne, Backus, Murnane, and Kapteyn}}]{Rundquist1998}
\bibinfo{author}{\bibfnamefont{A.}~\bibnamefont{Rundquist}},
  \bibinfo{author}{\bibfnamefont{C.~G.} \bibnamefont{Durfee}},
  \bibinfo{author}{\bibfnamefont{Z.}~\bibnamefont{Chang}},
  \bibinfo{author}{\bibfnamefont{C.}~\bibnamefont{Herne}},
  \bibinfo{author}{\bibfnamefont{S.}~\bibnamefont{Backus}},
  \bibinfo{author}{\bibfnamefont{M.~M.} \bibnamefont{Murnane}},
  \bibnamefont{and} \bibinfo{author}{\bibfnamefont{H.~C.}
  \bibnamefont{Kapteyn}}, \bibinfo{journal}{Science}
  \textbf{\bibinfo{volume}{280}}, \bibinfo{pages}{1412} (\bibinfo{year}{1998}).

\bibitem[{\citenamefont{Bartels et~al.}(2000)\citenamefont{Bartels, Backus,
  Zeek, Misoguti, Vdovin, Christov, Murnane, Kapteyn et~al.}}]{Bartels2000}
\bibinfo{author}{\bibfnamefont{R.}~\bibnamefont{Bartels}},
  \bibinfo{author}{\bibfnamefont{S.}~\bibnamefont{Backus}},
  \bibinfo{author}{\bibfnamefont{E.}~\bibnamefont{Zeek}},
  \bibinfo{author}{\bibfnamefont{L.}~\bibnamefont{Misoguti}},
  \bibinfo{author}{\bibfnamefont{G.}~\bibnamefont{Vdovin}},
  \bibinfo{author}{\bibfnamefont{I.}~\bibnamefont{Christov}},
  \bibinfo{author}{\bibfnamefont{M.}~\bibnamefont{Murnane}},
  \bibinfo{author}{\bibfnamefont{H.}~\bibnamefont{Kapteyn}},
  \bibnamefont{et~al.}, \bibinfo{journal}{Nature}
  \textbf{\bibinfo{volume}{406}}, \bibinfo{pages}{164} (\bibinfo{year}{2000}).

\bibitem[{\citenamefont{Drescher et~al.}(2001)\citenamefont{Drescher,
  Hentschel, Kienberger, Tempea, Spielmann, Reider, Corkum, and
  Krausz}}]{Drescher2001}
\bibinfo{author}{\bibfnamefont{M.}~\bibnamefont{Drescher}},
  \bibinfo{author}{\bibfnamefont{M.}~\bibnamefont{Hentschel}},
  \bibinfo{author}{\bibfnamefont{R.}~\bibnamefont{Kienberger}},
  \bibinfo{author}{\bibfnamefont{G.}~\bibnamefont{Tempea}},
  \bibinfo{author}{\bibfnamefont{C.}~\bibnamefont{Spielmann}},
  \bibinfo{author}{\bibfnamefont{G.~A.} \bibnamefont{Reider}},
  \bibinfo{author}{\bibfnamefont{P.~B.} \bibnamefont{Corkum}},
  \bibnamefont{and} \bibinfo{author}{\bibfnamefont{F.}~\bibnamefont{Krausz}},
  \bibinfo{journal}{Science} \textbf{\bibinfo{volume}{291}},
  \bibinfo{pages}{1923} (\bibinfo{year}{2001}).

\bibitem[{\citenamefont{Hentschel et~al.}(2001)\citenamefont{Hentschel,
  Kienberger, Spielmann, Reider, Milosevic, Brabec, Corkum, Heinzmann,
  Drescher, and Krausz}}]{Hentschel2001}
\bibinfo{author}{\bibfnamefont{M.}~\bibnamefont{Hentschel}},
  \bibinfo{author}{\bibfnamefont{R.}~\bibnamefont{Kienberger}},
  \bibinfo{author}{\bibfnamefont{C.}~\bibnamefont{Spielmann}},
  \bibinfo{author}{\bibfnamefont{G.}~\bibnamefont{Reider}},
  \bibinfo{author}{\bibfnamefont{N.}~\bibnamefont{Milosevic}},
  \bibinfo{author}{\bibfnamefont{T.}~\bibnamefont{Brabec}},
  \bibinfo{author}{\bibfnamefont{P.}~\bibnamefont{Corkum}},
  \bibinfo{author}{\bibfnamefont{U.}~\bibnamefont{Heinzmann}},
  \bibinfo{author}{\bibfnamefont{M.}~\bibnamefont{Drescher}}, \bibnamefont{and}
  \bibinfo{author}{\bibfnamefont{F.}~\bibnamefont{Krausz}},
  \bibinfo{journal}{Nature} \textbf{\bibinfo{volume}{414}},
  \bibinfo{pages}{509} (\bibinfo{year}{2001}).

\bibitem[{\citenamefont{Sansone et~al.}(2006)\citenamefont{Sansone, Benedetti,
  Calegari, Vozzi, Avaldi, Flammini, Poletto, Villoresi, Altucci, Velotta
  et~al.}}]{Sansone2006}
\bibinfo{author}{\bibfnamefont{G.}~\bibnamefont{Sansone}},
  \bibinfo{author}{\bibfnamefont{E.}~\bibnamefont{Benedetti}},
  \bibinfo{author}{\bibfnamefont{F.}~\bibnamefont{Calegari}},
  \bibinfo{author}{\bibfnamefont{C.}~\bibnamefont{Vozzi}},
  \bibinfo{author}{\bibfnamefont{L.}~\bibnamefont{Avaldi}},
  \bibinfo{author}{\bibfnamefont{R.}~\bibnamefont{Flammini}},
  \bibinfo{author}{\bibfnamefont{L.}~\bibnamefont{Poletto}},
  \bibinfo{author}{\bibfnamefont{P.}~\bibnamefont{Villoresi}},
  \bibinfo{author}{\bibfnamefont{C.}~\bibnamefont{Altucci}},
  \bibinfo{author}{\bibfnamefont{R.}~\bibnamefont{Velotta}},
  \bibnamefont{et~al.}, \bibinfo{journal}{Science}
  \textbf{\bibinfo{volume}{314}}, \bibinfo{pages}{443} (\bibinfo{year}{2006}).

\bibitem[{\citenamefont{Baker et~al.}(2006)\citenamefont{Baker, Robinson,
  Haworth, Teng, Smith, Chiril{\u{a}}, Lein, Tisch, and Marangos}}]{Baker2006}
\bibinfo{author}{\bibfnamefont{S.}~\bibnamefont{Baker}},
  \bibinfo{author}{\bibfnamefont{J.~S.} \bibnamefont{Robinson}},
  \bibinfo{author}{\bibfnamefont{C.~A.} \bibnamefont{Haworth}},
  \bibinfo{author}{\bibfnamefont{H.}~\bibnamefont{Teng}},
  \bibinfo{author}{\bibfnamefont{R.~A.} \bibnamefont{Smith}},
  \bibinfo{author}{\bibfnamefont{C.~C.} \bibnamefont{Chiril{\u{a}}}},
  \bibinfo{author}{\bibfnamefont{M.}~\bibnamefont{Lein}},
  \bibinfo{author}{\bibfnamefont{J.~W.~G.} \bibnamefont{Tisch}},
  \bibnamefont{and} \bibinfo{author}{\bibfnamefont{J.~P.}
  \bibnamefont{Marangos}}, \bibinfo{journal}{Science}
  \textbf{\bibinfo{volume}{312}}, \bibinfo{pages}{424} (\bibinfo{year}{2006}).

\bibitem[{\citenamefont{Baker et~al.}(2008)\citenamefont{Baker, Robinson, Lein,
  Chiril\ifmmode~\u{a}\else \u{a}\fi{}, Torres, Bandulet, Comtois, Kieffer,
  Villeneuve, Tisch et~al.}}]{Baker2008}
\bibinfo{author}{\bibfnamefont{S.}~\bibnamefont{Baker}},
  \bibinfo{author}{\bibfnamefont{J.~S.} \bibnamefont{Robinson}},
  \bibinfo{author}{\bibfnamefont{M.}~\bibnamefont{Lein}},
  \bibinfo{author}{\bibfnamefont{C.~C.} \bibnamefont{Chiril\ifmmode~\u{a}\else
  \u{a}\fi{}}}, \bibinfo{author}{\bibfnamefont{R.}~\bibnamefont{Torres}},
  \bibinfo{author}{\bibfnamefont{H.~C.} \bibnamefont{Bandulet}},
  \bibinfo{author}{\bibfnamefont{D.}~\bibnamefont{Comtois}},
  \bibinfo{author}{\bibfnamefont{J.~C.} \bibnamefont{Kieffer}},
  \bibinfo{author}{\bibfnamefont{D.~M.} \bibnamefont{Villeneuve}},
  \bibinfo{author}{\bibfnamefont{J.~W.~G.} \bibnamefont{Tisch}},
  \bibnamefont{et~al.}, \bibinfo{journal}{Phys. Rev. Lett.}
  \textbf{\bibinfo{volume}{101}}, \bibinfo{pages}{053901}
  (\bibinfo{year}{2008}).

\bibitem[{\citenamefont{Li et~al.}(2008)\citenamefont{Li, Zhou, Lock,
  Patchkovskii, Stolow, Kapteyn, and Murnane}}]{li2008}
\bibinfo{author}{\bibfnamefont{W.}~\bibnamefont{Li}},
  \bibinfo{author}{\bibfnamefont{X.}~\bibnamefont{Zhou}},
  \bibinfo{author}{\bibfnamefont{R.}~\bibnamefont{Lock}},
  \bibinfo{author}{\bibfnamefont{S.}~\bibnamefont{Patchkovskii}},
  \bibinfo{author}{\bibfnamefont{A.}~\bibnamefont{Stolow}},
  \bibinfo{author}{\bibfnamefont{H.~C.} \bibnamefont{Kapteyn}},
  \bibnamefont{and} \bibinfo{author}{\bibfnamefont{M.~M.}
  \bibnamefont{Murnane}}, \bibinfo{journal}{Science}
  \textbf{\bibinfo{volume}{322}}, \bibinfo{pages}{1207} (\bibinfo{year}{2008}).

\bibitem[{\citenamefont{Zhou et~al.}(2008)\citenamefont{Zhou, Lock, Li, Wagner,
  Murnane, and Kapteyn}}]{Zhou2008}
\bibinfo{author}{\bibfnamefont{X.}~\bibnamefont{Zhou}},
  \bibinfo{author}{\bibfnamefont{R.}~\bibnamefont{Lock}},
  \bibinfo{author}{\bibfnamefont{W.}~\bibnamefont{Li}},
  \bibinfo{author}{\bibfnamefont{N.}~\bibnamefont{Wagner}},
  \bibinfo{author}{\bibfnamefont{M.~M.} \bibnamefont{Murnane}},
  \bibnamefont{and} \bibinfo{author}{\bibfnamefont{H.~C.}
  \bibnamefont{Kapteyn}}, \bibinfo{journal}{Phys. Rev. Lett.}
  \textbf{\bibinfo{volume}{100}}, \bibinfo{pages}{073902}
  (\bibinfo{year}{2008}).

\bibitem[{\citenamefont{Itatani et~al.}(2004)\citenamefont{Itatani, Levesque,
  Zeidler, Niikura, P{\'e}pin, Kieffer, Corkum, and Villeneuve}}]{Itatani2004}
\bibinfo{author}{\bibfnamefont{J.}~\bibnamefont{Itatani}},
  \bibinfo{author}{\bibfnamefont{J.}~\bibnamefont{Levesque}},
  \bibinfo{author}{\bibfnamefont{D.}~\bibnamefont{Zeidler}},
  \bibinfo{author}{\bibfnamefont{H.}~\bibnamefont{Niikura}},
  \bibinfo{author}{\bibfnamefont{H.}~\bibnamefont{P{\'e}pin}},
  \bibinfo{author}{\bibfnamefont{J.~C.} \bibnamefont{Kieffer}},
  \bibinfo{author}{\bibfnamefont{P.~B.} \bibnamefont{Corkum}},
  \bibnamefont{and} \bibinfo{author}{\bibfnamefont{D.~M.}
  \bibnamefont{Villeneuve}}, \bibinfo{journal}{Nature}
  \textbf{\bibinfo{volume}{432}}, \bibinfo{pages}{867} (\bibinfo{year}{2004}).

\bibitem[{\citenamefont{Patchkovskii et~al.}(2006)\citenamefont{Patchkovskii,
  Zhao, Brabec, and Villeneuve}}]{Patchkovskii2006}
\bibinfo{author}{\bibfnamefont{S.}~\bibnamefont{Patchkovskii}},
  \bibinfo{author}{\bibfnamefont{Z.}~\bibnamefont{Zhao}},
  \bibinfo{author}{\bibfnamefont{T.}~\bibnamefont{Brabec}}, \bibnamefont{and}
  \bibinfo{author}{\bibfnamefont{D.~M.} \bibnamefont{Villeneuve}},
  \bibinfo{journal}{Phys. Rev. Lett.} \textbf{\bibinfo{volume}{97}},
  \bibinfo{pages}{123003} (\bibinfo{year}{2006}).

\bibitem[{\citenamefont{Patchkovskii et~al.}(2007)\citenamefont{Patchkovskii,
  Zhao, Brabec, and Villeneuve}}]{Patchkovskii2007}
\bibinfo{author}{\bibfnamefont{S.}~\bibnamefont{Patchkovskii}},
  \bibinfo{author}{\bibfnamefont{Z.}~\bibnamefont{Zhao}},
  \bibinfo{author}{\bibfnamefont{T.}~\bibnamefont{Brabec}}, \bibnamefont{and}
  \bibinfo{author}{\bibfnamefont{D.}~\bibnamefont{Villeneuve}},
  \bibinfo{journal}{J. Chem. Phys.} \textbf{\bibinfo{volume}{126}},
  \bibinfo{pages}{114306} (\bibinfo{year}{2007}).

\bibitem[{\citenamefont{Haessler et~al.}(2010)\citenamefont{Haessler, Caillat,
  Boutu, Giovanetti-Teixeira, Ruchon, Auguste, Diveki, Breger, Maquet,
  Carr{\'e} et~al.}}]{Haessler2010}
\bibinfo{author}{\bibfnamefont{S.}~\bibnamefont{Haessler}},
  \bibinfo{author}{\bibfnamefont{J.}~\bibnamefont{Caillat}},
  \bibinfo{author}{\bibfnamefont{W.}~\bibnamefont{Boutu}},
  \bibinfo{author}{\bibfnamefont{C.}~\bibnamefont{Giovanetti-Teixeira}},
  \bibinfo{author}{\bibfnamefont{T.}~\bibnamefont{Ruchon}},
  \bibinfo{author}{\bibfnamefont{T.}~\bibnamefont{Auguste}},
  \bibinfo{author}{\bibfnamefont{Z.}~\bibnamefont{Diveki}},
  \bibinfo{author}{\bibfnamefont{P.}~\bibnamefont{Breger}},
  \bibinfo{author}{\bibfnamefont{A.}~\bibnamefont{Maquet}},
  \bibinfo{author}{\bibfnamefont{B.}~\bibnamefont{Carr{\'e}}},
  \bibnamefont{et~al.}, \bibinfo{journal}{Nature Physics}
  \textbf{\bibinfo{volume}{6}}, \bibinfo{pages}{200} (\bibinfo{year}{2010}).

\bibitem[{\citenamefont{Vozzi et~al.}(2011)\citenamefont{Vozzi, Negro,
  Calegari, Sansone, Nisoli, De~Silvestri, and Stagira}}]{Vozzi2011}
\bibinfo{author}{\bibfnamefont{C.}~\bibnamefont{Vozzi}},
  \bibinfo{author}{\bibfnamefont{M.}~\bibnamefont{Negro}},
  \bibinfo{author}{\bibfnamefont{F.}~\bibnamefont{Calegari}},
  \bibinfo{author}{\bibfnamefont{G.}~\bibnamefont{Sansone}},
  \bibinfo{author}{\bibfnamefont{M.}~\bibnamefont{Nisoli}},
  \bibinfo{author}{\bibfnamefont{S.}~\bibnamefont{De~Silvestri}},
  \bibnamefont{and} \bibinfo{author}{\bibfnamefont{S.}~\bibnamefont{Stagira}},
  \bibinfo{journal}{Nature Physics} \textbf{\bibinfo{volume}{7}},
  \bibinfo{pages}{822} (\bibinfo{year}{2011}).

\bibitem[{\citenamefont{Lein et~al.}(2002{\natexlab{a}})\citenamefont{Lein,
  Hay, Velotta, Marangos, and Knight}}]{LeinPRL2002}
\bibinfo{author}{\bibfnamefont{M.}~\bibnamefont{Lein}},
  \bibinfo{author}{\bibfnamefont{N.}~\bibnamefont{Hay}},
  \bibinfo{author}{\bibfnamefont{R.}~\bibnamefont{Velotta}},
  \bibinfo{author}{\bibfnamefont{J.~P.} \bibnamefont{Marangos}},
  \bibnamefont{and} \bibinfo{author}{\bibfnamefont{P.~L.}
  \bibnamefont{Knight}}, \bibinfo{journal}{Phys. Rev. Lett.}
  \textbf{\bibinfo{volume}{88}}, \bibinfo{pages}{183903}
  (\bibinfo{year}{2002}{\natexlab{a}}).

\bibitem[{\citenamefont{Lein et~al.}(2002{\natexlab{b}})\citenamefont{Lein,
  Hay, Velotta, Marangos, and Knight}}]{LeinPRA2002}
\bibinfo{author}{\bibfnamefont{M.}~\bibnamefont{Lein}},
  \bibinfo{author}{\bibfnamefont{N.}~\bibnamefont{Hay}},
  \bibinfo{author}{\bibfnamefont{R.}~\bibnamefont{Velotta}},
  \bibinfo{author}{\bibfnamefont{J.~P.} \bibnamefont{Marangos}},
  \bibnamefont{and} \bibinfo{author}{\bibfnamefont{P.~L.}
  \bibnamefont{Knight}}, \bibinfo{journal}{Phys. Rev. A}
  \textbf{\bibinfo{volume}{66}}, \bibinfo{pages}{023805}
  (\bibinfo{year}{2002}{\natexlab{b}}).

\bibitem[{\citenamefont{Kanai et~al.}(2005)\citenamefont{Kanai, Minemoto, and
  Sakai}}]{Kanai2005}
\bibinfo{author}{\bibfnamefont{T.}~\bibnamefont{Kanai}},
  \bibinfo{author}{\bibfnamefont{S.}~\bibnamefont{Minemoto}}, \bibnamefont{and}
  \bibinfo{author}{\bibfnamefont{H.}~\bibnamefont{Sakai}},
  \bibinfo{journal}{Nature} \textbf{\bibinfo{volume}{435}},
  \bibinfo{pages}{470} (\bibinfo{year}{2005}).

\bibitem[{\citenamefont{Vozzi et~al.}(2005)\citenamefont{Vozzi, Calegari,
  Benedetti, Caumes, Sansone, Stagira, Nisoli, Torres, Heesel, Kajumba
  et~al.}}]{Vozzi2005}
\bibinfo{author}{\bibfnamefont{C.}~\bibnamefont{Vozzi}},
  \bibinfo{author}{\bibfnamefont{F.}~\bibnamefont{Calegari}},
  \bibinfo{author}{\bibfnamefont{E.}~\bibnamefont{Benedetti}},
  \bibinfo{author}{\bibfnamefont{J.-P.} \bibnamefont{Caumes}},
  \bibinfo{author}{\bibfnamefont{G.}~\bibnamefont{Sansone}},
  \bibinfo{author}{\bibfnamefont{S.}~\bibnamefont{Stagira}},
  \bibinfo{author}{\bibfnamefont{M.}~\bibnamefont{Nisoli}},
  \bibinfo{author}{\bibfnamefont{R.}~\bibnamefont{Torres}},
  \bibinfo{author}{\bibfnamefont{E.}~\bibnamefont{Heesel}},
  \bibinfo{author}{\bibfnamefont{N.}~\bibnamefont{Kajumba}},
  \bibnamefont{et~al.}, \bibinfo{journal}{Phys. Rev. Lett.}
  \textbf{\bibinfo{volume}{95}}, \bibinfo{pages}{153902}
  (\bibinfo{year}{2005}).

\bibitem[{\citenamefont{L'Huillier and Balcou}(1993)}]{L'Huillier1993}
\bibinfo{author}{\bibfnamefont{A.}~\bibnamefont{L'Huillier}} \bibnamefont{and}
  \bibinfo{author}{\bibfnamefont{P.}~\bibnamefont{Balcou}},
  \bibinfo{journal}{Phys. Rev. Lett.} \textbf{\bibinfo{volume}{70}},
  \bibinfo{pages}{774} (\bibinfo{year}{1993}).

\bibitem[{\citenamefont{Farrell et~al.}(2009)\citenamefont{Farrell, McFarland,
  G{\"u}hr, and Bucksbaum}}]{Farrell2009}
\bibinfo{author}{\bibfnamefont{J.}~\bibnamefont{Farrell}},
  \bibinfo{author}{\bibfnamefont{B.}~\bibnamefont{McFarland}},
  \bibinfo{author}{\bibfnamefont{M.}~\bibnamefont{G{\"u}hr}}, \bibnamefont{and}
  \bibinfo{author}{\bibfnamefont{P.}~\bibnamefont{Bucksbaum}},
  \bibinfo{journal}{Chem. Phys.} \textbf{\bibinfo{volume}{366}},
  \bibinfo{pages}{15 } (\bibinfo{year}{2009}).

\bibitem[{\citenamefont{McFarland et~al.}(2009)\citenamefont{McFarland,
  Farrell, Bucksbaum, and G\"uhr}}]{McFarland2009}
\bibinfo{author}{\bibfnamefont{B.~K.} \bibnamefont{McFarland}},
  \bibinfo{author}{\bibfnamefont{J.~P.} \bibnamefont{Farrell}},
  \bibinfo{author}{\bibfnamefont{P.~H.} \bibnamefont{Bucksbaum}},
  \bibnamefont{and} \bibinfo{author}{\bibfnamefont{M.}~\bibnamefont{G\"uhr}},
  \bibinfo{journal}{Phys. Rev. A} \textbf{\bibinfo{volume}{80}},
  \bibinfo{pages}{033412} (\bibinfo{year}{2009}).

\bibitem[{\citenamefont{Higuet et~al.}(2011)\citenamefont{Higuet, Ruf, Thir\'e,
  Cireasa, Constant, Cormier, Descamps, M\'evel, Petit, Pons
  et~al.}}]{Higuet2011}
\bibinfo{author}{\bibfnamefont{J.}~\bibnamefont{Higuet}},
  \bibinfo{author}{\bibfnamefont{H.}~\bibnamefont{Ruf}},
  \bibinfo{author}{\bibfnamefont{N.}~\bibnamefont{Thir\'e}},
  \bibinfo{author}{\bibfnamefont{R.}~\bibnamefont{Cireasa}},
  \bibinfo{author}{\bibfnamefont{E.}~\bibnamefont{Constant}},
  \bibinfo{author}{\bibfnamefont{E.}~\bibnamefont{Cormier}},
  \bibinfo{author}{\bibfnamefont{D.}~\bibnamefont{Descamps}},
  \bibinfo{author}{\bibfnamefont{E.}~\bibnamefont{M\'evel}},
  \bibinfo{author}{\bibfnamefont{S.}~\bibnamefont{Petit}},
  \bibinfo{author}{\bibfnamefont{B.}~\bibnamefont{Pons}}, \bibnamefont{et~al.},
  \bibinfo{journal}{Phys. Rev. A} \textbf{\bibinfo{volume}{83}},
  \bibinfo{pages}{053401} (\bibinfo{year}{2011}).

\bibitem[{\citenamefont{Bertrand et~al.}(2012)\citenamefont{Bertrand, W\"orner,
  Hockett, Villeneuve, and Corkum}}]{BertrandPRL2012}
\bibinfo{author}{\bibfnamefont{J.~B.} \bibnamefont{Bertrand}},
  \bibinfo{author}{\bibfnamefont{H.~J.} \bibnamefont{W\"orner}},
  \bibinfo{author}{\bibfnamefont{P.}~\bibnamefont{Hockett}},
  \bibinfo{author}{\bibfnamefont{D.~M.} \bibnamefont{Villeneuve}},
  \bibnamefont{and} \bibinfo{author}{\bibfnamefont{P.~B.}
  \bibnamefont{Corkum}}, \bibinfo{journal}{Phys. Rev. Lett.}
  \textbf{\bibinfo{volume}{109}}, \bibinfo{pages}{143001}
  (\bibinfo{year}{2012}).

\bibitem[{\citenamefont{Smirnova et~al.}(2009)\citenamefont{Smirnova, Mairesse,
  Patchkovskii, Dudovich, Villeneuve, Corkum, and Ivanov}}]{Smirnova2009}
\bibinfo{author}{\bibfnamefont{O.}~\bibnamefont{Smirnova}},
  \bibinfo{author}{\bibfnamefont{Y.}~\bibnamefont{Mairesse}},
  \bibinfo{author}{\bibfnamefont{S.}~\bibnamefont{Patchkovskii}},
  \bibinfo{author}{\bibfnamefont{N.}~\bibnamefont{Dudovich}},
  \bibinfo{author}{\bibfnamefont{D.}~\bibnamefont{Villeneuve}},
  \bibinfo{author}{\bibfnamefont{P.}~\bibnamefont{Corkum}}, \bibnamefont{and}
  \bibinfo{author}{\bibfnamefont{M.}~\bibnamefont{Ivanov}},
  \bibinfo{journal}{Nature} \textbf{\bibinfo{volume}{460}},
  \bibinfo{pages}{972} (\bibinfo{year}{2009}).

\bibitem[{\citenamefont{Fowe and Bandrauk}(2010)}]{Fowe2010}
\bibinfo{author}{\bibfnamefont{E.~P.} \bibnamefont{Fowe}} \bibnamefont{and}
  \bibinfo{author}{\bibfnamefont{A.~D.} \bibnamefont{Bandrauk}},
  \bibinfo{journal}{Phys. Rev. A} \textbf{\bibinfo{volume}{81}},
  \bibinfo{pages}{023411} (\bibinfo{year}{2010}).

\bibitem[{\citenamefont{W\"orner et~al.}(2010)\citenamefont{W\"orner, Bertrand,
  Hockett, Corkum, and Villeneuve}}]{WornerPRL2010}
\bibinfo{author}{\bibfnamefont{H.~J.} \bibnamefont{W\"orner}},
  \bibinfo{author}{\bibfnamefont{J.~B.} \bibnamefont{Bertrand}},
  \bibinfo{author}{\bibfnamefont{P.}~\bibnamefont{Hockett}},
  \bibinfo{author}{\bibfnamefont{P.~B.} \bibnamefont{Corkum}},
  \bibnamefont{and} \bibinfo{author}{\bibfnamefont{D.~M.}
  \bibnamefont{Villeneuve}}, \bibinfo{journal}{Phys. Rev. Lett.}
  \textbf{\bibinfo{volume}{104}}, \bibinfo{pages}{233904}
  (\bibinfo{year}{2010}).

\bibitem[{\citenamefont{Jin et~al.}(2012)\citenamefont{Jin, Bertrand, Lucchese,
  W\"orner, Corkum, Villeneuve, Le, and Lin}}]{Jin2012}
\bibinfo{author}{\bibfnamefont{C.}~\bibnamefont{Jin}},
  \bibinfo{author}{\bibfnamefont{J.~B.} \bibnamefont{Bertrand}},
  \bibinfo{author}{\bibfnamefont{R.~R.} \bibnamefont{Lucchese}},
  \bibinfo{author}{\bibfnamefont{H.~J.} \bibnamefont{W\"orner}},
  \bibinfo{author}{\bibfnamefont{P.~B.} \bibnamefont{Corkum}},
  \bibinfo{author}{\bibfnamefont{D.~M.} \bibnamefont{Villeneuve}},
  \bibinfo{author}{\bibfnamefont{A.-T.} \bibnamefont{Le}}, \bibnamefont{and}
  \bibinfo{author}{\bibfnamefont{C.~D.} \bibnamefont{Lin}},
  \bibinfo{journal}{Phys. Rev. A} \textbf{\bibinfo{volume}{85}},
  \bibinfo{pages}{013405} (\bibinfo{year}{2012}).

\bibitem[{\citenamefont{Zimmermann et~al.}(2005)\citenamefont{Zimmermann, Lein,
  and Rost}}]{Zimmermann2005}
\bibinfo{author}{\bibfnamefont{B.}~\bibnamefont{Zimmermann}},
  \bibinfo{author}{\bibfnamefont{M.}~\bibnamefont{Lein}}, \bibnamefont{and}
  \bibinfo{author}{\bibfnamefont{J.~M.} \bibnamefont{Rost}},
  \bibinfo{journal}{Phys. Rev. A} \textbf{\bibinfo{volume}{71}},
  \bibinfo{pages}{033401} (\bibinfo{year}{2005}).

\bibitem[{\citenamefont{Od\ifmmode~\check{z}\else \v{z}\fi{}ak and Milo\ifmmode
  \check{s}\else \v{s}\fi{}evi\ifmmode~\acute{c}\else
  \'{c}\fi{}}(2009)}]{Odzak2009}
\bibinfo{author}{\bibfnamefont{S.}~\bibnamefont{Od\ifmmode~\check{z}\else
  \v{z}\fi{}ak}} \bibnamefont{and} \bibinfo{author}{\bibfnamefont{D.~B.}
  \bibnamefont{Milo\ifmmode \check{s}\else \v{s}\fi{}evi\ifmmode~\acute{c}\else
  \'{c}\fi{}}}, \bibinfo{journal}{Phys. Rev. A} \textbf{\bibinfo{volume}{79}},
  \bibinfo{pages}{023414} (\bibinfo{year}{2009}).

\bibitem[{\citenamefont{Mairesse et~al.}(2008)\citenamefont{Mairesse, Levesque,
  Dudovich, Corkum, and Villeneuve}}]{Mairesse2008}
\bibinfo{author}{\bibfnamefont{Y.}~\bibnamefont{Mairesse}},
  \bibinfo{author}{\bibfnamefont{J.}~\bibnamefont{Levesque}},
  \bibinfo{author}{\bibfnamefont{N.}~\bibnamefont{Dudovich}},
  \bibinfo{author}{\bibfnamefont{P.~B.} \bibnamefont{Corkum}},
  \bibnamefont{and} \bibinfo{author}{\bibfnamefont{D.~M.}
  \bibnamefont{Villeneuve}}, \bibinfo{journal}{Journal of Modern Optics}
  \textbf{\bibinfo{volume}{55}}, \bibinfo{pages}{2591} (\bibinfo{year}{2008}).

\bibitem[{\citenamefont{Lewenstein et~al.}(1994)\citenamefont{Lewenstein,
  Balcou, Ivanov, L'Huillier, and Corkum}}]{Lewenstein1994}
\bibinfo{author}{\bibfnamefont{M.}~\bibnamefont{Lewenstein}},
  \bibinfo{author}{\bibfnamefont{P.}~\bibnamefont{Balcou}},
  \bibinfo{author}{\bibfnamefont{M.~Y.} \bibnamefont{Ivanov}},
  \bibinfo{author}{\bibfnamefont{A.}~\bibnamefont{L'Huillier}},
  \bibnamefont{and} \bibinfo{author}{\bibfnamefont{P.~B.}
  \bibnamefont{Corkum}}, \bibinfo{journal}{Phys. Rev. A}
  \textbf{\bibinfo{volume}{49}}, \bibinfo{pages}{2117} (\bibinfo{year}{1994}).

\bibitem[{\citenamefont{Etches and Madsen}(2010)}]{Etches2010}
\bibinfo{author}{\bibfnamefont{A.}~\bibnamefont{Etches}} \bibnamefont{and}
  \bibinfo{author}{\bibfnamefont{L.~B.} \bibnamefont{Madsen}},
  \bibinfo{journal}{J. Phys. B} \textbf{\bibinfo{volume}{43}},
  \bibinfo{pages}{155602} (\bibinfo{year}{2010}).
  
  \bibitem{jan}
  J. C. Baggesen and L. B. Madsen, J. Phys. B {\bf 44}, 115601 (2011).

\bibitem[{\citenamefont{Schmidt et~al.}(1993)\citenamefont{Schmidt, Baldridge,
  Boatz, Elbert, Gordon, Jensen, Koseki, Matsunaga, Nguyen, Su
  et~al.}}]{gamess}
\bibinfo{author}{\bibfnamefont{M.~W.} \bibnamefont{Schmidt}},
  \bibinfo{author}{\bibfnamefont{K.~K.} \bibnamefont{Baldridge}},
  \bibinfo{author}{\bibfnamefont{J.~A.} \bibnamefont{Boatz}},
  \bibinfo{author}{\bibfnamefont{S.~T.} \bibnamefont{Elbert}},
  \bibinfo{author}{\bibfnamefont{M.~S.} \bibnamefont{Gordon}},
  \bibinfo{author}{\bibfnamefont{J.~H.} \bibnamefont{Jensen}},
  \bibinfo{author}{\bibfnamefont{S.}~\bibnamefont{Koseki}},
  \bibinfo{author}{\bibfnamefont{N.}~\bibnamefont{Matsunaga}},
  \bibinfo{author}{\bibfnamefont{K.~A.} \bibnamefont{Nguyen}},
  \bibinfo{author}{\bibfnamefont{S.}~\bibnamefont{Su}}, \bibnamefont{et~al.},
  \bibinfo{journal}{J. Comput. Chem.} \textbf{\bibinfo{volume}{14}},
  \bibinfo{pages}{1347} (\bibinfo{year}{1993}).

\bibitem[{\citenamefont{Dunning}(1989)}]{Dunning1989}
\bibinfo{author}{\bibfnamefont{T.~H.} \bibnamefont{Dunning},
  \bibfnamefont{Jr.}}, \bibinfo{journal}{J. Chem. Phys.}
  \textbf{\bibinfo{volume}{90}}, \bibinfo{pages}{1007} (\bibinfo{year}{1989}).

\bibitem[{\citenamefont{Kendall et~al.}(1992)\citenamefont{Kendall, Dunning,
  and Harrison}}]{Dunning1992}
\bibinfo{author}{\bibfnamefont{R.~A.} \bibnamefont{Kendall}},
  \bibinfo{author}{\bibfnamefont{T.~H.} \bibnamefont{Dunning},
  \bibfnamefont{Jr.}}, \bibnamefont{and}
  \bibinfo{author}{\bibfnamefont{R.}~\bibnamefont{Harrison}},
  \bibinfo{journal}{J. Chem. Phys.} \textbf{\bibinfo{volume}{96}},
  \bibinfo{pages}{6796} (\bibinfo{year}{1992}).

\bibitem[{\citenamefont{Kono et~al.}(2001)\citenamefont{Kono, Koseki, Shiota,
  and Fujimura}}]{Kono2001}
\bibinfo{author}{\bibfnamefont{H.}~\bibnamefont{Kono}},
  \bibinfo{author}{\bibfnamefont{S.}~\bibnamefont{Koseki}},
  \bibinfo{author}{\bibfnamefont{M.}~\bibnamefont{Shiota}}, \bibnamefont{and}
  \bibinfo{author}{\bibfnamefont{Y.}~\bibnamefont{Fujimura}},
  \bibinfo{journal}{J. Phys. Chem. A} \textbf{\bibinfo{volume}{105}},
  \bibinfo{pages}{5627} (\bibinfo{year}{2001}).
  
  \bibitem{madsen}
  L. B. Madsen, Phys. Rev. A {\bf 65}, 053417 (2002).
  

\bibitem[{\citenamefont{Stapelfeldt and Seideman}(2003)}]{stapelfeldt_review}
\bibinfo{author}{\bibfnamefont{H.}~\bibnamefont{Stapelfeldt}} \bibnamefont{and}
  \bibinfo{author}{\bibfnamefont{T.}~\bibnamefont{Seideman}},
  \bibinfo{journal}{Rev. Mod. Phys.} \textbf{\bibinfo{volume}{75}},
  \bibinfo{pages}{543} (\bibinfo{year}{2003}).

\bibitem[{\citenamefont{W\"orner et~al.}(2009)\citenamefont{W\"orner, Niikura,
  Bertrand, Corkum, and Villeneuve}}]{WornerPRL2009}
\bibinfo{author}{\bibfnamefont{H.~J.} \bibnamefont{W\"orner}},
  \bibinfo{author}{\bibfnamefont{H.}~\bibnamefont{Niikura}},
  \bibinfo{author}{\bibfnamefont{J.~B.} \bibnamefont{Bertrand}},
  \bibinfo{author}{\bibfnamefont{P.~B.} \bibnamefont{Corkum}},
  \bibnamefont{and} \bibinfo{author}{\bibfnamefont{D.~M.}
  \bibnamefont{Villeneuve}}, \bibinfo{journal}{Phys. Rev. Lett.}
  \textbf{\bibinfo{volume}{102}}, \bibinfo{pages}{103901}
  (\bibinfo{year}{2009}).

\bibitem[{\citenamefont{Etches et~al.}(2011)\citenamefont{Etches, Gaarde, and
  Madsen}}]{Etches2011}
\bibinfo{author}{\bibfnamefont{A.}~\bibnamefont{Etches}},
  \bibinfo{author}{\bibfnamefont{M.~B.} \bibnamefont{Gaarde}},
  \bibnamefont{and} \bibinfo{author}{\bibfnamefont{L.~B.}
  \bibnamefont{Madsen}}, \bibinfo{journal}{Phys. Rev. A}
  \textbf{\bibinfo{volume}{84}}, \bibinfo{pages}{023418}
  (\bibinfo{year}{2011}).

\bibitem[{\citenamefont{Antoine et~al.}(1995)\citenamefont{Antoine, Piraux, and
  Maquet}}]{tp1}
\bibinfo{author}{\bibfnamefont{P.}~\bibnamefont{Antoine}},
  \bibinfo{author}{\bibfnamefont{B.}~\bibnamefont{Piraux}}, \bibnamefont{and}
  \bibinfo{author}{\bibfnamefont{A.}~\bibnamefont{Maquet}},
  \bibinfo{journal}{Phys. Rev. A} \textbf{\bibinfo{volume}{51}},
  \bibinfo{pages}{R1750} (\bibinfo{year}{1995}).

\bibitem[{\citenamefont{Kamta and Bandrauk}(2004)}]{tp2}
\bibinfo{author}{\bibfnamefont{G.}~\bibnamefont{Kamta}} \bibnamefont{and}
  \bibinfo{author}{\bibfnamefont{A.~D.} \bibnamefont{Bandrauk}},
  \bibinfo{journal}{Phys. Rev. A} \textbf{\bibinfo{volume}{70}},
  \bibinfo{pages}{011404} (\bibinfo{year}{2004}).

\bibitem[{\citenamefont{Chiril\ifmmode~\u{a}\else \u{a}\fi{}
  et~al.}(2010)\citenamefont{Chiril\ifmmode~\u{a}\else \u{a}\fi{},
  Dreissigacker, van~der Zwan, and Lein}}]{tp3}
\bibinfo{author}{\bibfnamefont{C.~C.} \bibnamefont{Chiril\ifmmode~\u{a}\else
  \u{a}\fi{}}},
  \bibinfo{author}{\bibfnamefont{I.}~\bibnamefont{Dreissigacker}},
  \bibinfo{author}{\bibfnamefont{E.~V.} \bibnamefont{van~der Zwan}},
  \bibnamefont{and} \bibinfo{author}{\bibfnamefont{M.}~\bibnamefont{Lein}},
  \bibinfo{journal}{Phys. Rev. A} \textbf{\bibinfo{volume}{81}},
  \bibinfo{pages}{033412} (\bibinfo{year}{2010}).

\bibitem[{\citenamefont{Gaarde et~al.}(2008)\citenamefont{Gaarde, Tate, and
  Schafer}}]{gaarde}
\bibinfo{author}{\bibfnamefont{M.~B.} \bibnamefont{Gaarde}},
  \bibinfo{author}{\bibfnamefont{J.~L.} \bibnamefont{Tate}}, \bibnamefont{and}
  \bibinfo{author}{\bibfnamefont{K.~J.} \bibnamefont{Schafer}},
  \bibinfo{journal}{J. Phys. B} \textbf{\bibinfo{volume}{41}},
  \bibinfo{pages}{132001} (\bibinfo{year}{2008}).

\bibitem[{\citenamefont{Antoine et~al.}(1996)\citenamefont{Antoine, L'Huillier,
  and Lewenstein}}]{antoine}
\bibinfo{author}{\bibfnamefont{P.}~\bibnamefont{Antoine}},
  \bibinfo{author}{\bibfnamefont{A.}~\bibnamefont{L'Huillier}},
  \bibnamefont{and}
  \bibinfo{author}{\bibfnamefont{M.}~\bibnamefont{Lewenstein}},
  \bibinfo{journal}{Phys. Rev. Lett.} \textbf{\bibinfo{volume}{77}},
  \bibinfo{pages}{1234} (\bibinfo{year}{1996}).

\bibitem[{\citenamefont{Bellini et~al.}(1998)\citenamefont{Bellini, Lyng\aa{},
  Tozzi, Gaarde, H\"ansch, L'Huillier, and Wahlstr\"om}}]{bellini}
\bibinfo{author}{\bibfnamefont{M.}~\bibnamefont{Bellini}},
  \bibinfo{author}{\bibfnamefont{C.}~\bibnamefont{Lyng\aa{}}},
  \bibinfo{author}{\bibfnamefont{A.}~\bibnamefont{Tozzi}},
  \bibinfo{author}{\bibfnamefont{M.~B.} \bibnamefont{Gaarde}},
  \bibinfo{author}{\bibfnamefont{T.~W.} \bibnamefont{H\"ansch}},
  \bibinfo{author}{\bibfnamefont{A.}~\bibnamefont{L'Huillier}},
  \bibnamefont{and} \bibinfo{author}{\bibfnamefont{C.-G.}
  \bibnamefont{Wahlstr\"om}}, \bibinfo{journal}{Phys. Rev. Lett.}
  \textbf{\bibinfo{volume}{81}}, \bibinfo{pages}{297} (\bibinfo{year}{1998}).

\bibitem[{\citenamefont{van~der Zwan et~al.}(2008)\citenamefont{van~der Zwan,
  Chiril\ifmmode~\u{a}\else \u{a}\fi{}, and Lein}}]{Zwan2008}
\bibinfo{author}{\bibfnamefont{E.~V.} \bibnamefont{van~der Zwan}},
  \bibinfo{author}{\bibfnamefont{C.~C.} \bibnamefont{Chiril\ifmmode~\u{a}\else
  \u{a}\fi{}}}, \bibnamefont{and}
  \bibinfo{author}{\bibfnamefont{M.}~\bibnamefont{Lein}},
  \bibinfo{journal}{Phys. Rev. A} \textbf{\bibinfo{volume}{78}},
  \bibinfo{pages}{033410} (\bibinfo{year}{2008}).

\bibitem[{\citenamefont{Keldysh}(1964) [Soc. Phys. -JETP {\bf 20}, 1307
  (1965)])}]{Keldysh}
\bibinfo{author}{\bibfnamefont{L.~V.} \bibnamefont{Keldysh}},
  \bibinfo{journal}{Zh. Eksp. Teor. Fiz.} \textbf{\bibinfo{volume}{47}},
  \bibinfo{pages}{1945} (\bibinfo{year}{1964) [Soc. Phys. -JETP {\bf 20}, 1307
  (1965)]}).

\bibitem[{\citenamefont{Faisal}(1973)}]{faisal}
\bibinfo{author}{\bibfnamefont{F.~H.~M.} \bibnamefont{Faisal}},
  \bibinfo{journal}{J. Phys. B: At. Mol. Phys.} \textbf{\bibinfo{volume}{6}},
  \bibinfo{pages}{L89} (\bibinfo{year}{1973}).

\bibitem[{\citenamefont{Reiss}(1980)}]{reiss}
\bibinfo{author}{\bibfnamefont{H.~R.} \bibnamefont{Reiss}},
  \bibinfo{journal}{Phys. Rev. A} \textbf{\bibinfo{volume}{22}},
  \bibinfo{pages}{1786} (\bibinfo{year}{1980}).

\end{thebibliography}

\end{document}